\documentclass[%
preprint,
showkeys,
 amsmath,amssymb,
 aps,
 prb,
]{revtex4-1}

\usepackage{graphicx}
\usepackage{dcolumn}
\usepackage{bm}

\begin{document}


\title{Pursuit of Wisdom and Quantum Ontology}

\author{Peter Kleinert}
\homepage{http://homepage.alice.de/p.kl/}
 \email{kl@pdi-berlin.de}


\date{\today}

\begin{abstract}
In his late work (De venatione sapientiae), Cusanus unfolded basic ideas of his brilliant theology. After a long period, this ingenious teaching became clearly recognizable especially in our time. Forward with his face to the back, modern scientific theory adopts nowadays a course to which Cusanus had already pointed centuries ago. Modern thought revolves with unexpected precision and unexpected mysteriousness around two issues of his doctrine of wisdom: (i) The possibility-of-being-made is not a figment of the human brain by which it organizes one's thoughts, but a fundamental and indispensable manifestation of reality. (ii) The possibility-of-being-made refers to something antecedent by which both the feasibility and the being-made get their common shape. This ultimate ground embodies the omnipotent oneness in the form of an infinite fund in which the cause of all reality and of all possibility is timelessly stored. Comparisons with the quantum ontology and the theory of quantum gravity impose themselves.
\end{abstract}

\maketitle


\section{\label{Intro}Introduction}
De venatione sapientiae is a late work of Nicholas of Kues (1401-1464), who is also known under the name Nicolaus Cusanus. With this scripture, he invites the reader to follow him in the hunt for wisdom. Accordingly, he shows the hunting ground, speculates where one can catch a lot of booty, sneaks up to wild animals, and finally presents his prey. The most eminent objective for his pursuit of wisdom was to gain an understanding of the given by means of its relationship to the giver. This concern corresponded in his day to the spirit of the time. What could not be doubted in the Middle Ages, was the belief that the ultimate truth of all beings rests in God. In line with this holy belief, Cusanus compared the universe with an organism, in which all parts cooperate under a divine plan. Without this divine providence, which has its origin in the absolute organizer, nothing can exist. Consequently, the hunter never could kill God, because otherwise neither hunting nor hunters would have existed.

During the centuries, the seemingly consistent medieval metaphysics eroded more and more. Let us mention only a few stages represented by influential opponents. First, there is the pioneer and founder of positivism in sociology, the French mathematician and philosopher Comte, who designated each religion as a childhood disease at an early stage in the history of mind, while Feuerbach unmasked the faith as self-delusion. But that was not enough, there's more. According to Marx, religion was nothing more than opium for people and Freud speculated that any religion has its root in infantile longing, which feels itself safe in the bosom of an omnipotent father. Finally, according to Nietzsche, God has died after a long and agonizing illness. The gesture of the believer who believes in the marvelousness was for the "{}philosopher with the hammer"{} beyond imagination. The most painful attack, however, came from the mechanistic philosophy of nature, which inexorably went from triumph to triumph. This attitude reached its culmination in the physical atomism, which came up at the beginning of the last century. According to this very successful scientific theory there are indestructible, eternal building blocks of matter, which can arrange in different ways to create a specific appearance of the world at each moment. To look for a mysterious substratum on which this mindless game with building blocks is founded proves to be completely pointless. Within a giant world machine there is no place for God. This opinion has become so hardened that even the soul is reduced to the world of things. The mechanistic world outlook is the natural philosophy of everyday activities, where each moment contributes to a history, which is once and for all being present with all its details, being fixed and in retrospect unchangeable. This experience is so overwhelming that religious speculations are only credible if they refer to the netherworld. In this context the hunt for wisdom originally initiated by Cusanus can no longer be accepted as an invitation to participate, but it is an old story, about which one can only gossip. The content of the analysis developed by one of the most deepest thinker of the world in his famous text seems already outdated. Thereby, Cusanus finished his hunting expedition with the hope that all his captured wisdom serves "{}for one's better speculating on these lofty matters."{}\cite{Cus1} However, for a long time scientists and philosophers were convinced that there is no kill in the hunting ground shown by Cusanus. In contrast to this opinion, let us ask: Is this disparaging assessment still valid? Our conclusion in this paper is amazing: The main ideas captured by Cusanus in his pursuit of wisdom can be regarded as a theological speculation which is consistent with modern quantum ontology that has been proven to be valid in countless experiments all over the world. De venatione sapientiae is therefore not only a document with reference to other old historical writings, but also a valuable proposal which serves as a promising starting point when trying to extend fundamental scientific knowledge into theological visions.

\section{\label{MdG}About the possibility-of-being-made}
\subsection{\label{E1}The hunting ground shown by Cusanus}
Cusanus founded his studies of wisdom on a firm basis, because he was convinced that "{}which is unknown cannot be known through that which is even more unknown."{}\footnote{Cf. Hopkins (1998), p. 6} Therefore, he started from scratch by capturing "{}something that is most certain - something presupposed and undoubted by all pursuers [of wisdom]$\dots$"{}\footnote{Cf. Hopkins (1998), p. 6} This self-evident base is a simple tautology in his scripture: Nothing will be done that cannot be done. From this comprehensible diagnosis, he concluded that the possibility-of-being-made precedes everything that has ever been made. While this statement still sounds convincing, we cannot help to wonder about his next step in the reasoning, when he attributed to the feasibility an "{}ontological status"{}. However, the possibility-of-being-made cannot have been made so that its "{}existence"{} should have something heavenly in itself. Nevertheless, according to Cusanus this mode of existence is an essential, indispensable, genuine part of the whole world. In his hunt for wisdom, the "{}ontological status" of the feasibility shows him the secret path to the fruitful game reserve, namely to something like the absolute beginning, "{}which is the Beginning and Cause of the possibility-of-being-made."{}\footnote{Cf. Hopkins (1998), p. 7} This preexistence cannot be made, because it precedes the possibility-of-being-made. Therefore, it is an unchangeable eternal source of all possibilities and realities. Just at this step, we arrive at the central idea of his teaching of wisdom: The Creator, who precedes the possibility-of-being-made as well as all existence "{}is the absolute and incontractible Beginning, for it is all that can be."{}\footnote{Cf. Hopkins (1998), p. 7} As the feasibility is neither feasible nor destructible, its beginning has a special character insofar as it "{}has not been made but, nevertheless, has a beginning, we speak of it as created, for it does not presuppose anything, from which it exists, except its Creator."{}\footnote{Cf. Hopkins (1998), p. 7} The possibility to make the world either way has its root in the eternal mind of God, who creates both being and nonbeing. From the almighty God (who is exactly what Cusanus called the possibility-of-making), the possibility-of-being-made was created from nothing. However, this creation is peculiar in the sense that its beginning and its end are indistinguishable. Everything can change - not God. He precedes each difference also the difference between possibility and reality. Only He is the actualized possibility-of-making, because He is what He can be. Whatever exists due to the possibility-of-being-made, exists actually only because it imitates the actuality of the possibility-of-making, which is the ultimate existence. All that has been made and all that could have been made "{}is subsequent to its own Actuality, which is Eternity."{}\footnote{Cf. Hopkins (1998), p. 36} Both features of the entire world, namely all that really exists and all that really can exist, coalesce within the eternal Creator. All life is only a shadow of Eternal Life.

\subsection{\label{K1}There is still large quarry}
The everyday experience deals with a world of hard-hitting and unchangeable facts that step forward for a brief moment to enlarge the realm of the non-varying past and to prepare the terrain for future tangible events. This succession has apparently no beginning and no end. The development is largely regular so that there is a whole universe of scientific questions that should be answered. Whether it is the anatomy of ants or the formation of stars after a supernova - the field of scientific research is almost boundless. The tremendous knowledge itself that is collected and stored in different ways, has a curious ontological status, since it is neither temporal nor  palpable. This form of existence is not written in stone but nebulous. It would be an obvious assumption that the cognitive ability emerges somehow from the reality of objective facts within the human brain. However, this modern attitude breaks with a long theological tradition, for which the world of relentless facts was incomplete and in itself not consistent. Indeed, the mechanistic world outlook is a shortcut, because the possibility-of-being-made seems to exist somehow, however, without any real ontological status. A consequence of this mechanistic bias is that all transcendental reflections about the specific suchness of existence are stigmatized as pointless. Inadmissible questions of this kind are, for instance: Was the feasibility created? Is the world a creation or part of a multiverse, in which almost everything can happen? However, if we pursue wisdom, then our question has an entirely different character: Does a transcendental principle reign behind the facade of seizable facts so that people must reconsider, what reality actually means? Cusanus answered in the affirmative by referring to the possibility-of-being-made as well as the possibility-of-making and by delegating true existence exclusively to the creator. It is amazing that modern science itself returns to the very same path. Thereby, science has absolutely nothing in mind with transcendence, although one concedes willingly that the new ontology is puzzling. In fact, the possibility-of-being-made is the focus of quantum ontology (see, for instance, Ref. \onlinecite{Griff_Ont} and Appendix A). Possibilities are ubiquitous in quantum theory. Mysteriously they arrive at their destination solely in our completely atypical, quasi-classical quantum world. The existence of concrete facts is linked to conditions, which can be precisely identified in quantum physics. Thus, the understanding of the possibility-of-being-made as an essential part of reality has nowadays taken a definite shape, which is known in details namely by quantum physics. In addition, also the origin of the specific possibility-of-being-made in our quasi-classical world can be investigated from a scientific point of view, although the solution of this problem is extremely difficult. Within the framework of the conventional quantum mechanics, the study must take into account not only the quantum dynamical laws, but also the initial quantum state. From this analysis it becomes evident that the quasi-classical world is not based on itself. Rather, one has to accept some overriding principle (theologically said: something transcendental), which is timeless, full of unimaginable possibilities (namely the many conceivable quantum universes) so that it can explain, how the world of tangible facts could emerge by means of a suitable initial state (by a "{}free volitional decision"{}). In conclusion, we admit that the prey of wisdom, which Cusanus presented in his scripture, gives us considerable food for thought. All his fundamental ontological problems have become nowadays a subject of scientific research (of course only in a specific unilateral form). A theological upgrading of all these scientific ideas was already anticipated by Cusanus so that a recollection to him is highly recommended.

\section{\label{God}What can we know about God?}
\subsection{\label{G1}The faceless oneness}
All real things and phenomena are anticipated by the possibility-of-being-made, which points to an origin, which itself is not made, but by which the feasibility gets its contours. What precedes existence and possible existence should be free from any intrinsic difference, since it is the indivisible cause of diversification. This profound oneness cannot be explained exhaustively, because it is the definition of itself. Everything that is definable has its origin in it. This ultimate ground, which permits to define everything as well as itself is nothing else but the not-other or the one. Just like God, the absolute eternal oneness is generally shaped by that what it can be. The omnipotent one is the beginning and the end of the unqualified possibility-of-being-made, whose singular contraction determines the essence of all things that are actually made. The oneness itself "{}is not essence, since it is the Cause of essence, for essence is something caused by it."{}\footnote{Cf. Hopkins (1998), p. 115} As God allows the understanding of all phenomena, he himself cannot be understood entirely. Rather, everything that is understandable is due to the possibility-of-being-made and is a representation of the eternal oneness that precedes understandability. Consequently, nothing fundamental can be learned about the omnipotent one. The true nature of God remains eternally hidden, not because our power to understand will never be sufficient, but because there is simply nothing to recognize. God is not an object of perceptibility. The divine "{}knowledge"{} of God is gained through ignorance that is the prerequisite for the pursuit of wisdom. Nevertheless, we see God's order, which is a sign of his government. At least, we realize that "{}the Divine Mind creates all things and always harmonizes all things and is the indestructible Cause of the order and harmony of all things."{}\footnote{Cf. Hopkins (1998), p. 89} The confusing ambiguity of our holy wisdom, its helplessness, is not a deficiency but the logical abandonment of pretensions to decrypt the divine origin of all phenomena. Enlightened people, who are educated about the limitations of the conceivability, find peace in the certainty to participate in the order and intention of the universe. All phenomena are integrated in the eternal harmonic organism, in which both reality and potentiality have their origin and which is worthy of praise beyond all limits. This confidence gets its completion by the religious faith in Jesus, who promises immortality of the soul. The humanity in Jesus is not only the unification of the lower with the higher nature, the timeliness with eternity, but simply the humanity of the Creator. We seek wisdom to become immortal. However, no wisdom can free us from death - this is the wrong route to wisdom. True "{}wisdom will be wisdom through which that necessity of dying is made into a virtue and will be wisdom which becomes for us a sure and safe passage unto the resurrection of life. This [passage] happens only by the power of Jesus and only for those who remain steadfastly on His pathway."{}\footnote{Cf. Hopkins (1998), p. 96}

\subsection{\label{G2}A maximally unknown miraculous number}
Indeed, the omnipotent one is an odd construction. On the one hand, it is some kind of divine knowledge, the cause of the possibility-of-being-made and of all thoughts by which phenomena become understandable. On the other hand we must admit that the oneness itself is neither understandable nor rooted in this world. Hence, our thoughts about the ultimate truths peter out. We must recognize: The last answers, which our heart desires so much, cannot reach us in principle. Therefore, Cusanus preached the learned ignorance, which plays an essential role in his philosophy.

This central idea of his teachings can be well illustrated by an unusual, irrational number $\Omega$, which is precisely defined, but, nevertheless, not computable. The ominous Chaitin number $\Omega$ denotes the probability that a randomly generated string of bits proves to be a program that runs on a computer and eventually halts after a finite time. As the halting problem cannot always be decided in advance, this number is fundamentally not countable. Moreover, $\Omega$ is not only maximally uncomputable, but also maximally unknowable and maximally random (some further information is compiled in Appendix \ref{AppendixB}). What is most interesting for our purpose: Chaitin's number $\Omega$ "{}is also the diamond-hard distilled and crystallized essence of mathematical truth."{}\cite{CSam1} It is amazing: The answer to every mathematical question is written down in omega, even though, we basically cannot distill the universal mathematical wisdom from $\Omega$. Like a cabalistic number, the digits of $\Omega$ encode the secrets of the whole mathematical universe. Unfortunately, this digit sequence is always uncomputable so that we finally know nothing more about $\Omega$ than its maximal indefiniteness (its maximal randomness). Our talking about $\Omega$ is nothing more than learned ignorance. Only God knows whether a given randomly generated computer program will eventually halt or not. For Him, $\Omega$ is both palpable and understandable, just as the sum of all possible mathematical theorems. Somewhat boldly, we can assert: "{}$\dots$ if you believe in $\Omega$, then you believe in God,"{}\cite{CSam2} and vice versa. As according to modern quantum physics true randomness governs our quantum universe, we live in a world that is infinitely complex (like the incompressible random number $\Omega$) and, therefore, is unknowable in its entirety by intellectual beings. By this mathematical example, the learned ignorance, preached by Cusanus, becomes certainly more apperceptible.

\section{\label{Mind}On the origin of the possibility-of-being-made}
\subsection{Creation of feasibility}
The design by which the possibility-of-being-made is predetermined is not made, since it is in fact the foundation of feasibility. This eternal omnipotent design is all that it can be in as much as it is not itself a form, but the source of all forms, which determine what actually could happen via the possibility-of-being-made. But what can specify the variety of forms if not mind and wisdom? From the intrinsic predetermination of all forms by the divine mind, all things obtain their suchness by means of the delimitation. By creating the feasibility, the divine wisdom arranges the world in such a way as it was predefined by eternity. The possibility-of-being-made was created so that it produces this and no other world. The creation owes its occurrence to the free will of the divine mind, whose foundation is nothing else than spirit. This holy wisdom is the origin of all possible and real forms, which all have a natural affinity to God in accordance to their ancestry. The true nature, the genuine forms of all phenomena are exclusively present in God. Only due to His intellect all things exist. In contrast, the human intellect assimilates intelligent things and looks for his own understanding by constructing his own images. Employing the intelligent assimilation, people gain a vague insight of the world. However, the divine essence of all things and therefore also the real essence of rationality cannot be considered.

\subsection{Generalized quantum ontology}
Cusanu's ingenious speculation about the predetermination of the possibility-of-being-made finds a striking illustration by modern discussions aimed at generalizing quantum physics.

With respect to phenomena occurring in nature, the possibility-of-being-made is expressed by fundamental physical laws, which are all quantum mechanical, and which look for universal regularities in physical systems. Fundamental is the state vector, which is defined on a space-like Cauchy surface and which contains all information of the quantum system (a Cauchy surface is a three-dimensional surface in four-dimensional space-time with the property that no point is the future or past of a point of the very same surface). The quantum mechanical state vector shifts along a family of space-like surfaces via an unitary evolution or state reduction. Due to the one-parametrical foliation of Cauchy surfaces, the notions of time and history become extraordinarily essential in the conventional formulation of quantum physics. As mentioned in Appendix \ref{AppendixA}, consistency limits the prediction of the theory to probabilities of decoherent sets of alternative histories. The quasi-classical predictability, which is ubiquitous in our universe, is an emergent feature of the particular initial condition and the particular dynamical laws. Other sets of decohering histories may have the same source, but may differ, nevertheless, profoundly from the well acquainted quasi-classical realm. These uninhabitable alternatives correspond to complementary ways of speaking about the unfolding of the same initial condition.

Cusanus did not finish his pursuit of wisdom with this whole entangled conception of reality, possibility, and timeliness. Rather, he thought of something, what precedes timeliness so that he has to answer the difficult question: How can temporality originate without time? The same problem tantalizes modern theoretical physics, which gives a remarkable and interesting general answer, although important details are still unsettled. It turns out that the usual framework with unitarily evolving states on space-like surfaces is not the most fundamental formulation of quantum theory. In quantum gravity, the existence of a fixed background space-time geometry with a well defined causal structure is only ensured in special situations when quantum fluctuations of space and time are smooth in the vicinity of a saddle point of the Euclidean action functional.\cite{Hartle_Hawking} Space and time are notions that apply only to special circumstances in the much wider realm of quantum gravity so that space-time is in fact emergent. In general, quantum histories "{}do not have to represent evolution {\it in} space-time. Rather, they can be histories {\it of} space-time."{}\cite{Hartle2006} What the unitary quantum evolution on a fixed quasi-classical background geometry precedes is a timeless theory of quantum gravity, in which all the glory of the physical universe is once and for all present without any change and without any possibility to add or remove something. Scientists can approach this holy physical wisdom only step by step by designing appropriate pictures of the divine essence of the physical world. The ultimate knowledge is not deducible, because it is not knowledge as we know it but in a composite knowledge and the prerequisite of knowledge together, as Cusanus speculated in his famous hunt for wisdom. In addition, the sequence of the scientific progress points to a vision of the world, which is much more complex than deepest thinkers could ever imagine. What in this way vaguely shows up on the horizon that religious people always have called God.

\section{\label{Infinity}There is no plurality of infinity}
\subsection{\label{I1}Infinity and oneness}
In his study of equality, Cusanus clarified an issue which is important for the understanding of his conviction. He argued that nothing real existing is exactly repeatable so that there are no real phenomena all over the world which are precisely equal. Exact equality is not of this world, because it would be all what it can be. Therefore, equality is prior to inequality and has the status of eternity. By contrast, what "{}can be made more equal is subsequent to the possibility-of-being-made."{}\footnote{Cf. Hopkins (1998), p. 68} The equality in and of itself, which is only approximated by the equality of real things, cannot be increased, because it is once and for all being given in the same manner as all archetypical, eternal, primordial shapes. "{}For just as goodness, beauty, truth, etc., which in eternity are Eternity itself, are also so equal that they are Equality, which is Eternity: so they are not more than one. Likewise, there cannot be a plurality of eternal things, since the Eternal is Actualized-possibility, i.e., is [actually] that which unqualifiedly can be. And likewise all eternal things are not more than one eternal thing, even as eternal Goodness, eternal Greatness, eternal Beauty, eternal Truth, eternal Equality are not more than one eternal thing. Similarly, they are not a plurality of equal things, because they are so equal that they are most simple Equality itself, which precedes all plurality."{}\footnote{Cf. Hopkins (1998), p. 96} From the uniqueness of eternity, a fundamental existential dualism arises. On the one hand, there are finite images which all differ from each other and are therefore never that what they might be. On the other hand, there is the domain of eternity, in which we project different beings, although there exists only the omnipotent oneness, namely God. "{}But Equality itself is the Word of Not-other, i.e., of God the Creator, who defines and speaks of both Himself and all things."{}\footnote{Cf. Hopkins (1998), p. 70}

\subsection{\label{I2}About the heaven of infinities}
The infinite is not a biblical term, because appearently no positive assertion is attached to it. The infinite became important for the first time in the doctrine of God, which was proposed by Gregory of Nyssa (335-394). According to Gregory, infinity no longer meant formlessness and indefiniteness, but fullness and glory. Unfortunately, an infinite God cannot be perceived so that people cannot uplift on him. The resulting fatalistic piety, which was especially advocated by Dionysius Areopagita, leads to a dead end. Cusanus, who stood in the tradition of Dionysius, was able to avoid the crash into the fatalism of the mystic theology. Admittedly, also Cusanus was teaching the learned ignorance with great emphasis. But he was convinced that through an elevated talking, the preacher becomes aware of some ciphers of God.

The theology of Nicholas of Cusa implies an infinite God, who unfolds itself (via the possibility-of-being-made) and secretes itself (as the infinite oneness, which is beyond recognition) at the same time. Everything that is being made is already prefigured in the feasibility. The feasibility itself is an endless fund, which is partly reflected in the made and which occupies an ontological status even more than "{}reality"{}. Very impressively Cusanus illustrated his thoughts through a meditation about seeing (De visione Dei). His findings were summarized as follows: Seeing is at the same time a being seen. All the functional principles of seeing - physical, chemical, biological, and physiological - as well as all cognitive activities by which the vision fulfills its objective, are pre-formulated in the endless reservoir of possibilities of seeing. This fund, which extends to the Big Bang when the habitable universe was born, has its true counterpart in the actuality of the possibility-of-making (in God). Here, the authentic watching is found with all its infinite wealth, which cannot help but watches too, however, in an unprecedented manner. Our seeing is based on a holy seeing, which cannot be further perfected and which watches at us with the divine eye. This heavenly seeing is inexplorable as it is absorbed in the ethereal oneness, where all the divine idealities are united, namely in God.

The parallel to the quantum ontology is obvious. Above all, quantum physics deals with possibilities, in which all what is feasible is pre-formulated. The actually being made is the "{}realization"{} of these possibilities in a strange quantum universe, namely in our quasi-classical world. The seeing, the thinking, the feeling, etc. have an equivalent in the possibility-of-being-made (in quantum-mechanical probabilities), which cannot increase or decrease and which all together create, therefore, the holy unity (God).

For Cusanus, it was quite natural to assume that there could not be a hierarchy of infinities. Anything that cannot be bigger, smaller, better, smarter, etc. merge into the divine oneness without distinction. It is completely obvious that his vision of God would be destroyed by accepting a plurality of infinities. A heaven with infinitely many heavens was unthinkable. First Cantor (1845-1918), the founder of modern set theory, demonstrated the conceivability of an infinite hierarchy of infinities within mathematics. Appendix \ref{AppendixC} provides a brief overview of big and very big infinities in set theory. The construction of this hierarchy is so problematic that famous mathematicians strictly disagreed. Nevertheless, the adventurous ideas about hierarchical infinities appeared already in philosophy, for instance, in the philosophy of world ages by Schelling. It would be an intellectual adventure of particular kind to extend Nicholas of Cusa's theology so that it can handle infinities of different nature. Such ideas would certainly found their counter part in a speculative quantum ontology.

\section{Conclusion}
What Cusanus occupied in his day, namely the pursuit of wisdom, seems to be inexplicable for many contemporaries. Their world view is crucially shaped by the mechanistic materialism, which is encouraged by the glorious findings of classical physics. Reality means for those people the totality of all tangible things that change with time. Possibilities as well as the whole reality of spirit have no ontological status in this ideology. Perhaps, it is assumed that these phenomena are emergent oddities, which exist only in the mind of people. In any case, the strict atomism does not open any free room to integrate both potentiality and mind into a unified world view. Where Cusanus once went on the hunt, there are now allotments and tenements. A hunt can no longer be organized there. Is this gloomy picture the final counsel of the scientific progress that the hunt for wisdom is futile? Does our enlightenment ruin the hunt? Many signs speak for the conclusion that these questions must be answered with yes. To mention just an example: Numerous popular scientific writings (for instance: L. Susskind, "{}The Cosmic Landscape"{}, A. Vilenkin, "Many Worlds in One"{}, S. Lloyd, "{}Programming the Universe"{}) paint a grandiose picture of the world, whereby they distinctly disassociate themselves from wisdom (respectively from the glorious theological tradition). However, this ignorance will not endure. With respect to physical reality, modern theoretical physics step by step approaches ideas that find a counterpart in the inspired theology of Nicholas of Cusa and other famous theologians. We must realize: Reality is not only the material world, but at the same time also the possibility-of-being-made. Reality without possibility is incomprehensible. This more fundamental point of view concerns not only the future but also the past.\cite{Hartle1997} Moreover, the tangible reality is nothing more than a peculiar potentiality, which allows observations of compact entities in a very special quantum world. But it goes even further: Modern theoretical physics meets Cusanus in the conviction that the total world affairs of possibility and reality, of the possibility-of-being-made is not the final answer. There must be something that orchestrates the whole spectacle. And this source is neither finite nor temporal but infinite and timeless. The theory of quantum gravitation seeks to discover such a rigid frame, in which our transient world has its very special place. To speak of God in this scientific thoughts is obviously inappropriate. Nevertheless, we must recognize that the grandiose findings of modern physics find a much better theological appreciation by the wisdom of Cusanus than by the narrow-minded mechanistic materialism. The turn to the wisdom of the ancients is not only justified by modern scientific thinking, but also urgently required for several reasons.

\appendix

\section{\label{AppendixA}Consistent quantum theory}
In the course of the last century, an exciting new fundamental conception of the physical world was discovered that has been verified by numerous experiments. Strictly speaking, there are up to now no conflicting observations, which jeopardize the progress of quantum physics. This revolutionary development enforces a rigorous rethinking not only in sciences but also in philosophy and theology. In order to successfully promote this reformation, a consistent interpretation of quantum physics is indispensable. Recent publications (for instance Refs. [\onlinecite{Grif,Hoh}]) give an well understandable presentation of the matter. However, a reader without any interest in physics will have some trouble to accept many of the discussed unfamiliar thoughts. Since the reevaluation of Cusanu's pursuit of wisdom relies on a familiarity with the new material, namely the quantum ontology, let us illustrate the basic findings by a simple picture.%
\begin{figure}[htbp]
\includegraphics[width=0.39\textwidth]{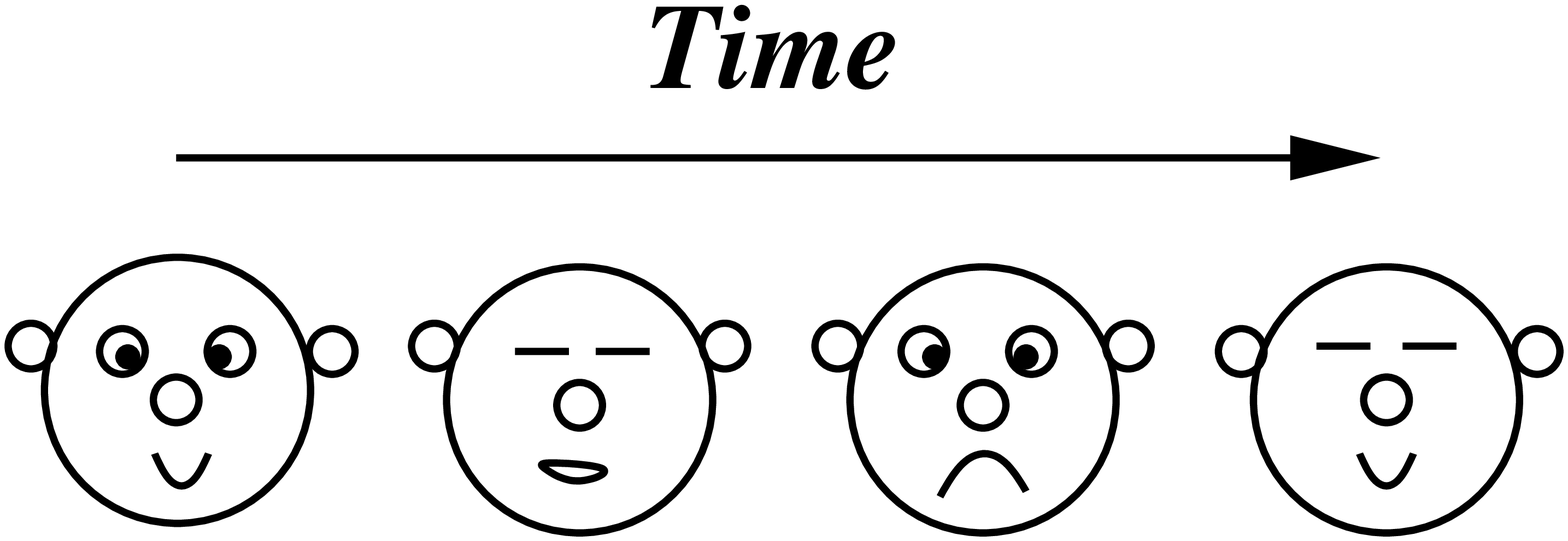}
\caption{\label{fratze1}In the quasi-classical world, all features of an object (face) are clearly observable at each moment. Consequently, in hindsight there is only one unique history.}
\end{figure}
The cartoon in Fig.~\ref{fratze1} shows a head with its characteristic features namely ears, eyes, mouth, and nose. According to classical physics and the common sense, at any moment of time, the face exhibits all its peculiarities so that it is everything that it can be at each instant. Consequently, the past is an unchangeable but non-existing "{}reality"{} and only the future is subject to the possibility-of-being-made. Hence, the origin of the feasibility remains fundamentally inexplicable and beyond of our tangible world. Based on this classical ontology, the hunt for wisdom proposed by Cusanus cannot be more than a daydream.

\begin{figure}[htbp]
\includegraphics[width=0.39\textwidth]{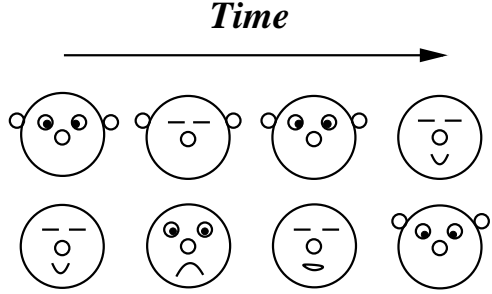}
\caption{\label{fratze2}The quantum face is different. In the quantum world there are always much more features of an object (face) than can in principle emerge at an instant. Consequently, the history is basically ambiguous.}
\end{figure}

It turned out, however, that reality is by nature completely different. To get an idea of what is the fundamental true face of reality, lets switch to Fig.~\ref{fratze2}, which illustrates the mysterious ghost story about the quantum world.%
It is essential to realize that the quantum face can never exhibit all its features at a given time. To give an example lets say that having a mouth and having ears is mutually incompatible so that it will never happen that we see a face exhibiting both a mouth and ears at a given time. According to this basic complementarity principle of quantum theory, there are always more features that a quantum object or phenomenon could possibly exhibit than those which actually constitute reality at any time. Consequently, from a strict quantum mechanical point of view, there is no thing in reality which is all that what it can be. The classicality is excluded by quantum mechanical complementarity. To understand the "{}reality"{} of a quantum object, we must account its possibility-of-being-made by determining the probabilities for its various possible manifestations. Strictly speaking, we are never able to say this or that will happen, but only this or that could happen. What is surprising, however, we live in a very special world in which decoherence permanently leads to the appearance of a classical universe with an extremely peaked probability distribution and a blurring of incompatibilities. Up to now there is no final theory by which one can conclusively explain the occurrence of the quasi-classical world. Nevertheless, there is no doubt that the quasi-classical realm of every day experience is due to the early quantum state of the universe together with the quantum dynamical laws.\cite{Hartle2008} It remains in force that the possibility-of-being-made has a strict ontological status in our world as it was claimed by Cusanus long ago in his famous pursuit of wisdom. In addition, we see from Fig.~\ref{fratze2} that there are always many different incompatible histories in quantum universes, a fact that has illuminating implications for the concept of time.

The following comments are closer to the seriousness of the well-funded scientific findings. The orthodox quantum physics finds its natural description by a Hilbert space $\cal{H}$, which is a linear complete metric vector space. In this abstract mathematical structure there are orthonormal sequences of basis vectors $|\psi_j\rangle$, which are complete so that the identity operator $I$ of $\cal{H}$ can be represented by orthogonal projections $P_j=|\psi_j\rangle\langle \psi_j|$ on the Hilbert space by means of the equation $I=\sum_{j}P_j$. The state vectors are eigenfunctions of the projection operators $P_j|\psi_i\rangle=\delta_{ij}|\psi_j\rangle$ with the eigenvalues $0$ or $1$. Therefore, the projectors $P_j$ have the function to decide whether the system is in the state $|\psi_j\rangle$ or not. However, the decomposition of the $I$-operator is not unique, since the set of basis vectors that span the Hilbert space is not unique. Two decompositions $I=\sum_{j}P_j$ and $I=\sum_{j}Q_j$ are called compatible if all projectors commute $[P_i,Q_j]=0$ for all $i,j$, because commuting hermitian operators have the same basis. What is most essential: in quantum theory, physical quantities are expressed by operators that do not necessarily commute with each other. Non-commuting operators, which have no equivalent in classical physics, give rise to incompatible representations, which should be strictly separated in order to avoid serious paradoxes. Consequently, what a physicist figures out about a given quantum system depends on his question. The evaluation of a given snapshot requires a fixing of the viewpoint, because there are always many different incompatible possibilities to think about the same thing, a situation which would be simply absurd in classical physics. What the quantum physicist is concerned with is nothing else than the possibility-of-being-made. He cannot say: this or that will happen, rather he has to specify his questioning (the specific representation for his study, which he decides to prefer compared to other incompatible possibilities) in order to get nothing more than probabilities for many things that {\it could} happen. The possibilities-of-being-made are not superficial conditions of existence which dictate the motion of divine, eternal elementary particles, but are intrinsically entangled with that what is called reality. In this general sense, we are really forced to agree with the texts of Cusanus.

The principle ambiguity of scenarios that refer to one and the same phenomenon at a given time has implications for the understanding of quantum histories, too. Since at any given time $t_i$ ($i=1,2\dots ,n$), a specific viewpoint (set of basis vectors) has to be chosen to answer meaningful questions about future evolutions, the Hilbert space of histories $\widetilde{\cal{H}}$ has to be a tensor product $\widetilde{\cal{H}}={\cal H}_{1}\otimes{\cal{H}}_{2}\otimes\dots{\cal{H}}_{n}$ where ${\cal H}_i$ is a copy of the Hilbert space that refers to the instant $t_i$. Now, we can repeat our above considerations to explain what a quantum history actually can be. As the projectors $P_{j}^{\alpha_j}$ of a temporary decomposition of the Hilbert space ${\cal H}_j$ at time $t_j$ form a complete set, we have $I_j=\sum_{\alpha_j}P_j^{\alpha_j}$. The tensor product of these projectors $Y^{\alpha}=P_{1}^{\alpha_1}\otimes\dots\otimes P_{n}^{\alpha_n}$ (with $\alpha =\{\alpha_j \}$) is a projector in the Hilbert space $\widetilde{\cal{H}}$, from which the identity operator $\widehat{I}$ in the space of histories can be constructed $\widehat{I}=\sum_{\alpha}Y^{\alpha}$. Again, the principle of complementarity comes into play, as the incompatibilities in the original Hilbert space ${\cal H}$ carries over to the Hilbert space $\widetilde{\cal{H}}$ of histories. That means, the decomposition $\widehat{I}=\sum_{\alpha}Y^{\alpha}$, which characterizes a viewpoint that allows the appreciation of each history $Y$ by exploiting $Y=\sum_{\alpha}c_{\alpha}Y^{\alpha}$ (with $c_{\alpha}$ being complex numbers), is not unique. There are always many incompatible possibilities to tell a story about the same thing. But what actually are inconsistent histories? To assign probabilities to possible events requires that all histories in question are mutually consistent so that the decoherence functional $D(\alpha^{\prime},\alpha)={\rm Tr}(Y^{\alpha^{\prime}}\rho {Y^{\alpha}}^{\dag})$ becomes diagonal. Only in this case (if $D(\alpha^{\prime},\alpha)=0$ for $\alpha^{\prime}\neq \alpha$), we are able to say something about current and future events. Under these circumstances, our research terminates with the conclusion: the history $Y^{\alpha}$ could possibly be happen with probability $p(\alpha)=D(\alpha,\alpha)$. But there are also incompatible histories $Y^{\alpha^{\prime}}$ and ${Y^{\alpha}}$ for which $D(\alpha^{\prime},\alpha)\neq 0$ holds true. All what we can say about the combination of such inconsistent histories is simply: {\it nonsense}. The principle variety of parallel histories is an undeniable fact discovered by modern physics, which, however, is completely intolerable for the common sense. Nevertheless, there are already imprints in the history of philosophy that came close to that findings (an example is the philosophy of world ages by Schelling).

The density matrix $\rho$, which enters the definition of the decoherence functional, is crucial for the determination of historical processes that could occur. The possibility-of-being-made with respect to histories is determined not only by equations of motion (from which the eigenstates $|\psi_j\rangle$ and therefore the projectors $P_j=|\psi_j\rangle\langle \psi_j|$ are calculated), but also by the "{}initial condition"{} expressed by the density matrix $\rho$. For the simple case that there was an initial pure quantum state $|\psi_0\rangle$, we have $\rho=|\psi_0\rangle\langle \psi_0|$. The determination of the initial condition is a serious and complicated problem. A self-consistent solution proposed by J. Hartle, S. Hawking, and T. Hertog is particularly promising.\cite{Hartle_Hawking}

\section{\label{AppendixB}The mysterious number $\Omega$}
Any scientific law, any information about technological processes, historical events, psychological abnormalities and so on can be expressed in terms of binary numbers. Therefore, the general study of strings, which digitally encode information, essentially contributes to the epistemology. It is surprising that these more formal analysis conveys deep insights into what cognition means. A famous example is Chaitin's number $\Omega$. The idea is pretty simple. Let us generate a long disordered binary string by chance and interpret the result as a random computer program. The probability to obtain a string with $n$ bits amounts $(1/2)^{n}$. Among the variety of randomly produced binary strings there are countless computer programs that could initiate real computations. The seemingly simple question is: Can we decide in a finite amount of time, whether or not the respective computer program consisting of $n$ bits ever halts? The general answer is: No. This counterintuitive finding, proved by Turing, allows the definition of a number that is not countable, namely the probability $\Omega$ that a randomly generated program ever halts. The formal expression for Chaitin's number is $\Omega=\sum_{p\, {\rm halts}}2^{-|p|}$, where $|p|$ represents the length of the computer program $p$ in bits. The irrational number $\Omega$ exhibits some remarkable peculiarities. Because $\Omega$ is based on an unsolvable problem, namely Turing's halting problem, its complexity is really infinite. It's infinitely many digits have a "{}divine origin"{}, because their determination would require infinite resources, which are not available. So $\Omega$ is really a very complex number. That means that the smallest computer program that generates $\Omega$ cannot be shorter than $\Omega$ itself (however, there is no algorithm which can decide, whether a program is as small as possible). As there is no redundancy in the digit stream of $\Omega$, there is also no mathematical recipe (no finite binary string), by which $\Omega$ could be calculated (as it is the case for other irrational numbers). The inventory of $\Omega$ is not preserved in an elegant theory, which is encoded by a finite binary sequence. The absolute complexity of $\Omega$ means complete lawlessness and pure  arbitrariness. About $\Omega$ one cannot speak in pictures and metaphors. This number is maximally unknowable and random. The best one could say about $\Omega$ would be a transmission of all its digits what is impossible.

\section{\label{AppendixC}Infinite sets}
The set of all rational numbers is infinite, but countable, whereas there is another infinity, namely the set of all real numbers, which has a greater "{}power"{}, because it is not countable due to the additional infinite (not countable) set of irrational numbers (an assertion, which was critizised recently, c.f., Ref. \onlinecite{Zenkin}). Consequently, not all infinite sets have the same "{}size"{}. But how can we distinguish infinite sets from each other? The famous mathematician, Georg Cantor (1845-1918), solved this problem by establishing modern set theory. General counting is not based on integers, but on well-ordered sets, which are totally ordered with the addition that every nonempty subset has a least member so that a unique successor can be identified in each infinite set. A well-ordered set ${\cal A}:\equiv\langle A,<_{A}\rangle$ encompasses both the finite or infinite set $A$ and the well-ordering relation $<_{A}$. According to the well-ordering theorem, each set can be well-ordered. This finding suggests the introduction of ordinal numbers $\alpha$, which are strictly well-ordered and transitive so that every element of $\alpha$ is also a subset of $\alpha$. It is the key feature of ordinals that each well-ordered set is order isomorph to exactly one ordinal number. Therefore, generalized counting can exploit the isomorphism between well-ordered sets $\langle A,<_{A}\rangle$ and $\langle \alpha,<_{\alpha}\rangle$ with an unique ordinal $\alpha$. There is an endless supply of ordinals, such that (i) one ordinal is the first, (ii) each ordinal has a successor, (iii) for each set of ordinals there is another ordinal, which succeeds them all. The collection $O_{n}$ of all ordinals is an inconsistent, absolutely infinite multiplicity. The smallest infinite ordinal number $\omega$ represents countable infinite sets. At the next step, we encounter the set of real numbers (the continuum) and the Euclidean space $\mathbb{R}$, the "{}measure"{} of which is the power set $\mathcal{P}(\omega)$ (the set of all subsets of $\omega$). All sets at this level are no longer countable in the conventional sense. Its binary representation is a function from $\omega$ to $\{0,1\}$, which are elements of the power set $\{0,1\}^{\omega}$. What sounds amazing, the equivalence relation $\mathbb{R}^{\omega}\sim\mathbb{R}$ holds true, which means that an arbitrarily short part of a straight line can be mapped onto an Euclidean space with countable infinite dimensions. An even bigger infinity accounts for all functions, which map $\mathbb{R}$ onto $\mathbb{R}$: $\mathbb{R}^{\mathbb{R}}\sim\mathcal{P}(\mathbb{R})\sim\mathcal{P}(\mathcal{P}(\omega)))$. Usually, sets with higher cardinality are not treated in mathematics. To proceed further in the hierarchy of infinities, lets introduce the notion of cardinals. Based on the axiom of choice, we identify the cardinal number with the initial ordinal. Cardinals are, therefore, special ordinals that are not equipotent to any smaller ordinal. Cardinal numbers, usually expressed by the aleph $\aleph$, are particularly suitable for counting. The set $Cn_{\infty}$ of all infinite cardinal numbers has as its smallest element $\aleph_{0}\sim\omega$ the cardinality of natural numbers. The aleph function generates an isomorphism between $\langle On,<_{On}\rangle$ and $\langle Cn_{\infty},<\rangle$ so that for each cardinal $\kappa$ there exists an index $\alpha$ with $\kappa=\aleph_{\alpha}$. What is astonishing, the set $Cr(Cn_{\infty})$ of all critical alephs, which satisfy the fixed-point relation $\kappa=\aleph_{\kappa}$, has the same size (cardinality) as the set $O_{n}$ of all ordinals. Moreover, if $A$ is an unbounded, closed set of ordinal numbers than the set $Cr(A)$ of all critical cardinals is also an infinite, unbounded and closed set of ordinals. Consequently, the universe of sets expands explosively by collecting all sets with critical cardinals: $\Theta_{0},\,\Theta_{1},\dots\Theta_{\omega},\,\Theta_{\omega+1},\dots\Theta_{\aleph_1},\dots\Theta_{\Theta_0},\dots$ It is a special feature of these uncountable, weakly inaccessible cardinals that they cannot be generated from below by formations of power sets. These inaccessible cardinals have such a large cardinality that they contain the entire classical set theory. The existence of inaccessible cardinals is not guaranteed by ordinary axioms of set theory. It is the universe axiom (or equivalently the inaccessible cardinal axiom) that ensures the existence of an infinite tower of inaccessible cardinals. And on top of the entire mountain reside the inconsistent, absolutely infinite sets, which Cantor identified with God. However, the idea of an actual infinity (formally proposed by the axiom of infinity) has attracted a lot of hostility, first of all by mathematical schools such as constructivism and intuitionism.


\providecommand{\noopsort}[1]{}\providecommand{\singleletter}[1]{#1}%

\end{document}